\documentclass{aa}
\usepackage{graphics,times}

\begin{document}

   \title{Starbursts and torus evolution in AGN}

   \authorrunning{Vollmer, Beckert, \& Davies}

   \author{B.~Vollmer\inst{1}, T.~Beckert\inst{2}, \and R.I.~Davies\inst{3}}

   \offprints{B.~Vollmer, e-mail: bvollmer@astro.u-strasbg.fr}

   \institute{CDS, Observatoire astronomique de Strasbourg, UMR 7550, 
              11, rue de l'universit\'e, 67000 Strasbourg, France \and
              Max Planck Institut f\"ur Radioastronomie, 
              Auf dem H\"ugel 69, 53121 Bonn, Germany, \and
              Max Planck Insitut f\"ur extraterrestrische Physik, Postfach 1312,
              85741, Garching, Germany}

   \date{Received / Accepted}

\abstract{
Recent VLT SINFONI observations of the close environments ($\sim 30$~pc) of nearby AGNs have shown 
that thick gas tori and starbursts with ages between $10$ and $150$~Myr are frequently found.
By applying these observations to a previously established analytical model of clumpy 
accretion disks, we suggest an evolutionary sequence for starburst and AGN phases.
Whereas the observed properties of the gas tell us about the current state of the torus,
the starburst characteristics provide information on the history of the torus.
In the suggested evolution, a torus passes through 3 different phases predetermined by 
an external mass accretion rate. Started by an initial, short, and massive gas infall,
a turbulent and stellar wind-driven $Q \sim 1$ disk is formed in which the starburst proceeds.
Once the supernovae explode the intercloud medium is removed, leaving a massive, geometrically thick,  
collisional disk
with a decreasing, but still high-mass accretion rate. When the mass accretion rate has significantly
decreased, the collisional torus becomes thin and transparent as the circumnuclear disk
in the Galactic center of the Milky Way.  
Variations on this scenario are possible either when there is a
second short and massive gas infall, in which case the torus may
switch back into the starburst mode, or when there is no initial short
massive gas infall. All observed tori up to now have been collisional and thick.
The observations show that this phase can last more than $100$~Myr. 
During this phase the decrease in the mass accretion rate within the torus 
is slow (a factor of 4 within $150$~Myr). The collisional tori also form stars, but with an efficiency
of about $10$\,\% when compared to a turbulent disk.
 }

\maketitle

\section{Introduction \label{sec:intro}}
In the unification scheme for active galactic nuclei (AGN) the central massive
black hole is surrounded by a geometrical thick gas and dust torus
(see, e.g., Antonucci 1993). If the
observer's line-of-sight crosses the torus material, the AGN is entirely
obscured from near-IR to soft X-rays and only visible at X-ray energies if the gas column density is not too high
(Seyfert 2 galaxies).
On the other hand, if the torus is oriented face-on with respect to the observer,
the central engine is visible (Seyfert 1 galaxies).
The spectral energy distributions (SEDs) of most quasars and AGN in Seyfert galaxies have a pronounced 
secondary peak in the mid-infrared (mid-IR) (e.g. Sanders et al. 1989; Elvis et al. 1994), which 
is interpreted as thermal emission by hot dust in the torus. The dust is heated by the primary 
optical/ultraviolet (UV) continuum radiation, and the torus extends from the dust sublimation 
radius outwards (Barvainis 1987). 

The geometrical thickness of the torus in the gravitational potential of the galactic nucleus
implies a vertical velocity dispersion of about
$50$-$100$~km\,s$^{-1}$. If one assumes that the disk is continuous, i.e. thermally supported,
this corresponds to a temperature of $\sim 10^5$~K. Since this is beyond the dust
sublimation temperature ($\sim 10^3$~K), thick tori have to be clumpy 
or must be supported by additional forces other than thermal pressure.
Krolik \& Begelman (1988) proposed a clumpy torus model where the clumps have supersonic velocities.
Vollmer et al. (2004) and Beckert \& Duschl (2004) elaborated this model
in which orbital motion can be randomized if magnetic fields permit the cloud collisions
to be sufficiently elastic. Vollmer et al. (2004) found that the
circumnuclear disk (CND) in the
Galactic center (G\"{u}sten et al. 1987) and obscuring tori share the same gas physics, 
where the mass of clouds is  in the range 20 - 50 $M_\odot$ and
their density close to the limit of disruption by tidal shear. 
A change in matter supply and the dissipation of kinetic energy can turn a torus into a
CND-like structure and vice versa.
Any massive torus will naturally lead to sufficiently high mass accretion
rates to feed a luminous AGN.

If and how efficient these clumpy tori form stars is an open question.
The large majority of observational studies probed the nuclear star formation 
on scales of a few hundred parsecs (see, e.g., Sarzi et al. 2007, Asari et al. 2007,
Gonz\'ales Delgado \& Cid Fernandes 2005, Cid Fernandes et al. 2004).
These studies resulted in a general view that about $30$\,\% - $50$\,\% of
the sample AGNs are associated with recent (ages less than a few 100~Myr) star formation
on these scales. Thanks to the high spatial resolution of the near-infrared adaptive optics
integral field spectrograph SINFONI, it has only recently become possible to study the
environments of AGN on the 10~pc scale. Davies et al. (2007) analyzed star formation
in the nuclei of nine Seyfert galaxies at spatial resolutions down to $0.085''$.
They found recent, but no longer active, starbursts in the central regions
which occurred $10$ - $300$~Myr ago. Moreover, 
Hicks et al. (2008) were able to measure
the rotation and dispersion velocity of the molecular gas 
in these galaxies using the $2.12$~$\mu$m 
H$_2$ (1-0)S(1) line. Surprisingly, all molecular gas tori have high velocity dispersions
and are therefore geometrically thick.

In this article we compare 
the observations of Davies et al. (2007) and Hicks et al. (2008)
with the expectations of 
analytical models of clumpy
accretion disks developed in Vollmer \& Beckert (2002, 2003) and Vollmer et al. (2004).
In a first step, we test 
whether these models are able to describe observations.
In a second step, these models allow us to investigate the scenario hypothesized by
Davies et al. (2007) where sporadic, short-lived
starbursts are due to short massive accretion events in the central region, 
followed by more quiescent phases until there is another episode of accretion.
This scenario is corroborated by a closer look at the nucleus of NGC~3227 
where Davies et al. (2006) found signs of a past starburst ($\sim 40$~Myr ago)
and a presently quiescent gas torus with a Toomre parameter $Q>1$.

\section{The theory of clumpy gas disks \label{sec:theory}}

Within the framework of Vollmer \& Beckert (2002, 2003) and Vollmer et al. (2004)
clumpy accretion disks are divided into two categories:
(i) turbulent and (ii) collisional disks. In case (i) the ISM is regarded as
a single entity which changes phase (molecular, atomic, ionized) according 
to internal (gas density, pressure, magnetic field) and external (gravitation,
radiation field, winds) conditions. Energy is injected into a turbulent cascade 
at the driving length scale (large scale) and dissipated
at the dissipation length scale (small scale). We identify the dissipation length scale with 
the characteristic size of selfgravitating clouds. 
These clouds decouple from the the turbulent cascade and constitute the first energy sink.
The source of energy which is injected at the driving scale to maintain turbulence
can be either (i) mass accretion in the gravitational
potential of the galactic center ({\it fully gravitational FG model})
or (ii) supernova explosions ({\it SN model}).
In the collisional case energy is also supplied in the process of mass accretion in the gravitational
potential of the galactic center and dissipated via partially
inelastic cloud--cloud collisions. The actual dissipation rate in individual collisions is largely unknown.
The disk evolution is mainly driven by the external mass accretion rate.
Since these models are equilibrium models, we assume that the mass accretion rate
is constant throughout the region of interest when averaged for a sufficiently long time 
($\sim \Omega^{-1}$; here $\Omega$ is the angular velocity of circular orbits in the gravitational 
potential of the galatic nucleus) and maintained for at least the turnover timescale $R/v_{\rm turb}$ 
of gas in the disk, where $R$ is the distance from the center of the galaxy and $v_{\rm turb}$ the 
characteristic speed of turbulent eddies.
All models give access to the global parameters of the disk 
and the local parameters of the most massive clouds (see Table~\ref{tab:model}).
The free parameters of the models are the Toomre parameter $Q$, the disk transparency $\vartheta$
(Eq.~\ref{eq:tau}), and the mass accretion rate $\dot{M}$. Other parameters are fixed using the Galactic
values (Vollmer \& Beckert 2002, 2003). Each model has an associated star formation rate.
In the following we describe these models in more detail.

\begin{table}
      \caption{Model parameters and their meaning}
         \label{tab:model}
      \[
       \begin{array}{|l|l|}
        \hline
	{\rm large\ scale} & {\rm disk} \\
	\hline
	\hline
	R & {\rm galactic\ radius} \\
	v_{\rm rot} & {\rm rotation\ velocity} \\
	\Omega & {\rm angular\ velocity} \\
	M_{\rm dyn} & {\rm total\ enclosed\ (dynamical)\ mass} \\
	M_{\rm gas} & {\rm total\ gas\ mass} \\
	v_{\rm turb} & {\rm gas\ turbulent\ velocity\ dispersion} \\
	Q & {\rm Toomre\ parameter} \\
	\rho & {\rm midplane\ density} \\
	\Sigma & {\rm surface\ density} \\
	H & {\rm disk\ height} \\
	\dot{M} & {\rm disk\ mass\ accretion\ rate} \\
	\nu & {\rm viscosity} \\
	l_{\rm driv} & {\rm turbulent\ driving\ length\ scale} \\
	\delta & {\rm scaling\ parameter\ between\ driving}\\
	  & {\rm scale\ length\ and\ cloud\ size}\\
	\Phi_{\rm V} & {\rm cloud\ volume\ filling\ factor} \\
	\zeta & {\rm viscosity\ scaling\ parameter} \\
	t_{\rm ff}^{\rm H} & {\rm disk\ vertical\ free\ fall\ time} \\
	\dot{\rho}_* & {\rm star\ formation\ rate\ per\ unit\ volume}\\
	\dot{\Sigma}_* & {\rm star\ formation\ rate\ per\ unit\ surface}\\
	\dot{M}_* & {\rm star\ formation\ rate} \\
	\xi & {\rm conversion\ factor\ for\ SN\ energy\ flux}\\
	\eta & {\rm star\ formation\ efficiency\ in\ the}\\
	  & {\rm collisional\ model}\\
	\gamma & {\rm linking\ factor\ between\ M_{\rm gas}\ and\ \dot{M}}\\
	  & {\rm for\ torus\ evolution}\\  
	\vartheta & {\rm disk\ transparency} \\
	\dot{M}_{\rm BH} & {\rm mass\ accretion\ rate\ onto\ the\ black\ hole}\\
	\dot{M}_{\rm wind} & {\rm wind\ outflow\ rate}\\
	J_{\rm UV} & {\rm AGN\ UV\ radiation\ field}\\
	\hline
	\hline
	{\rm small\ scale} & {\rm clouds}  \\
	\hline
	\hline
	t_{\rm coll} & {\rm timescale\ for\ cloud-cloud\ collisions}\\ 
	l_{\rm coll} & {\rm cloud\ mean\ free\ path} \\ 
	M_{\rm cl} & {\rm cloud\ mass} \\
	r_{\rm cl} & {\rm cloud\ radius} \\
	N_{\rm cl} & {\rm cloud\ surface\ density} \\
	c_{\rm s} & {\rm local\ sound\ speed\ within\ the\ clouds}\\
	t_{\rm s} & {\rm sound\ crossing\ time\ of\ clouds} \\
	t_{\rm ff}^{\rm cl} & {\rm cloud\ free\ fall\ time}\\
	T & {\rm temperature} \\
	c_{\rm i} & {\rm sound\ crossing\ time\ of\ the\ ionized\ gas}\\
	\hline
        \end{array}
      \]
\end{table}

\subsection{Turbulent disks}

In Vollmer \& Beckert 2002 (Paper I) we developed an analytical
model for clumpy accretion disks and included a simplified description 
of turbulence in the disk. In contrast to classical accretion disk theory 
(see, e.g., Pringle 1981),
we eliminated the  ``thermostat'' mechanism, which implies a direct 
coupling between the heat produced by viscous friction and the 
viscosity itself. The viscosity is usually assumed to be proportional
to the thermal sound speed. Thus, 
the (gas) heating rate depends itself on the gas temperature. 
This leads to an equilibrium corresponding to a thermostat mechanism.
Instead, we use energy flux conservation, where the potential
energy that is gained through mass accretion and differential rotation
is cascaded by turbulence from large to small scales and dissipated there. 

One fundamental approximation is that the kinetic energy is
dissipated (removed from the turbulent cascade) when the gas clouds become self-gravitating. 
Turbulence transfers the energy from the driving wavelength $l_{\rm driv}$ to the
dissipation wavelength $l_{\rm diss}$, which corresponds to the size of the largest selfgravitating clouds.
The two length scales are linked by the scaling parameter $\zeta$.
For a Kolmogorov-like turbulent energy spectrum
$\zeta = (l_{\rm driv}/l_{\rm diss})^{\frac{3}{4}}$. In addition, the 
modeled disks have a constant Toomre-$Q$ parameter:
\begin{equation}
Q = \frac{v_{\rm turb}\,\Omega}{\pi\,G\,\Sigma} \geq 1\ ,
\label{eq:tq}
\end{equation}
where $v_{\rm turb}$ is the turbulent velocity, $\Omega$ the angular
velocity,
$G$ the gravitational constant, and $\Sigma$ the gas surface density of 
the disk. 
If we can approximate the total gas mass within a radius $R$ by
$M_{\rm gas} = \pi R^2 \Sigma$ the Toomre parameter can be rewritten
\begin{equation}
Q=\frac{v_{\rm turb}}{v_{\rm rot,K}}\frac{M_{\rm dyn}}{M_{\rm gas}}\ ,
\label{eq:tqint}
\end{equation}
where $M_{\rm dyn}$ is the total enclosed mass and $v_{\rm rot}$ 
the Keplerian rotation velocity.

Furthermore, we use the following prescription for the viscosity:
\begin{equation}
\nu=\zeta^{-1} v_{\rm turb} l_{\rm driv}\ .
\label{eq:nuu}
\end{equation}
We obtained a set of equations with 3 free parameters:
the Toomre parameter $Q$, the scaling parameter $\zeta > 1$, and
the mass accretion rate within the disk $\dot{M}$. The mass accretion rate is the gas mass 
transported per time through the gas disk from large to small radii. It must be supplied from 
the galaxy at the outer radius of the turbulent gas disk and is constant at all radii in the disk.
This set of equations can be solved analytically and the results were 
already used in Paper I to describe properties of our Galaxy. 
The solutions depend
on the parameters $Q$, $\zeta$, $\dot{M}$, $v_{\rm rot}$, and $R$.
It turned out that the driving wavelength equals the disk height
$l_{\rm driv}=H$.

In a second step we included the energy input due to supernova (SN)
explosions (Vollmer \& Beckert 2003, Paper II). 
The energy flux provided by SNe is 
transfered by turbulence to smaller scales where it is again dissipated. 
The SN energy flux is assumed to be proportional 
to the local star formation rate. The local star formation 
rate $\dot{\rho}_{*}$ is taken to be proportional
to the mean density and inversely proportional to the local free fall
time of the clouds. These clouds have sizes that are a factor $\delta$
smaller than the driving length scale. 
The factor of proportionality is the probability to find
a self-gravitating cloud, i.e. the volume filling factor.
The integration length in the vertical $z$ direction is assumed to be the turbulent
driving scale length, i.e. the vertical height in the disk where self-gravitating clouds are found:
$\dot{\Sigma}_{*}=\dot{\rho}_{*}\,l_{\rm driv}$.
The SN energy per unit time $\dot{E}_{\rm SN}$ per area $\Delta A$ is
therefore proportional to the local star formation rate 
$\dot{\Sigma}_{*}$:
\begin{equation}
\frac{\dot{E}_{\rm SN}}{\Delta A} = \xi \dot{\Sigma}_{*}\ ,
\label{eq:sfrsn}
\end{equation}
where the factor of proportionality $\xi$ is independent of the radius 
in the disk.
Its normalization with Galactic values yields $\xi = 4.6\,10^{-8}$~(pc/yr)$^{2}$.

In the FG model the energy transported through the turbulent cascade is 
supplied by mass accretion, which leads to an energy flux equation of the form 
\begin{equation}
\rho \nu \frac{v_{\rm turb}^{2}}{l_{\rm driv}}=-\frac{1}{2\pi}\dot{M}
v_{\rm rot} \frac{\partial \Omega}{\partial R}\ .
\label{eq:efc}
\end{equation}
In the case of SN driven turbulence the energy flux is determined 
by the star formation rate
\begin{equation}
\rho \nu \frac{v_{\rm turb}^{2}}{l_{\rm driv}}=\xi\,\dot{\Sigma}_{*}\ .
\label{eq:sgzefc}
\end{equation}
Furthermore, we take into account that the clouds are formed during the interaction between SN remnants at the
compressed edges. The size of
the clouds $l_{\rm cl}$ is consequently smaller than the turbulent driving
wavelength and we use $\delta=l_{\rm driv}/l_{\rm cl}$ 
with $\delta \geq 1$.

\subsection{Collisional disks and the stability of the clouds \label{sec:colldisk}}

If the selfgravitating clouds are stable, their collisions will give rise to angular 
momentum redistribution again described by an effective viscosity. 
An equilibrium disk can be formed if there are fragmenting
collisions or partially elastic collisions (the clouds are supposed to be magnetized).
If the collisional timescale $t_{\rm coll}$ is longer or
equal to the dynamical timescale, the resulting viscosity can be written as 
\begin{equation} \label{eq:vis}
\nu = \vartheta^{-1}v_{\rm turb} H\ ,
\label{eq:tau}
\end{equation}
where the disk transparency $\vartheta = t_{\rm coll}\Omega > 1$ and $H$ is the disk height.
It follows that the collisional energy dissipation rate is 
\begin{equation}
\frac{\Delta E}{\Delta A\,\Delta t}=f \frac{\Sigma v_{\rm turb}^{2}}{t_{\rm coll}}=
f \frac{\Sigma v_{\rm turb}^{3}}{l_{\rm coll}}=f \frac{\Sigma v_{\rm turb}^{3}}{\vartheta H} .
\end{equation}
Here the factor $f$, which has been omitted in Vollmer et al. (2004),  accounts for the mean fraction 
of cloud mass participating in the highly supersonic cloud collisions. For constant density 
clouds Krolik \& Begelman (1988) argue that $f = 0.2$. For more centrally condensed, 
self-gravitating clouds we will use a factor $f=0.1$ in this paper. This geometric 
factor $f$ not only reduces the average energy dissipation rate in cloud collisions but 
also the angular momentum redistribution in the collisions. The reduced energy dissipation 
is  accompanied by a correspondingly reduced mass accretion rate $\dot{M}$ for the same 
transparency in the model disks $\nu \rightarrow f \nu$ in Eq.(\ref{eq:vis}).

Since in the FG model $l_{\rm driv}=H$, the FG model and the collisional model
are formally equivalent.
We can thus use Eq.~(\ref{eq:tq} - \ref{eq:nuu}) replacing $\zeta$ by $\vartheta$ to describe 
the collisional model.
Nonetheless the interpretation of $\zeta$ and $\vartheta$ is completely different.

The cloud size $r_{\rm cl}$ and the volume filling factor $\Phi_{\rm V}$ of clouds 
can be derived using their mean free path (see Vollmer et al. 2004) 
\begin{equation}
l_{\rm coll}=\vartheta H = \frac{4r_{\rm cl}}{3 \Phi_{\rm V}}
\end{equation}
and the fact that the clouds are selfgravitating
\begin{equation}
t^{\rm cl}_{\rm ff}=\sqrt{\frac{3 \pi \Phi_{\rm V}}{32 G \rho}}= t_{\rm s} = \frac{r_{\rm cl}}{c_{\rm s}}\ ,
\end{equation}
where $t^{\rm cl}_{\rm ff}$ is the free fall timescale within clouds, $\rho$ the disk overall gas density, 
$t_{\rm s}$ the sound crossing timescale, and $c_{\rm s}$ the sound speed.
This leads to
\begin{equation}
r_{\rm cl}=\frac{\pi^2}{8}\frac{Q c_{\rm s}^2}{\Omega v_{\rm turb} \vartheta}
\end{equation}
and
\begin{equation}
\Phi_{\rm V}=\frac{\pi^2}{6} \frac{Q c_{\rm s}^2}{v_{\rm turb}^2 \vartheta^2}\ .
\end{equation}
The cloud mass is then
\begin{equation}
M_{\rm cl}=\frac{4 \pi}{3} \Phi_{\rm V}^{-1} \rho r_{\rm cl}^3=\frac{2^{\frac{1}{3}}\pi^4}{64} c_{\rm s}^4
\vartheta^{-\frac{4}{3}} G^{-\frac{4}{3}} Q^{\frac{2}{3}} f^{\frac{1}{3}} \dot{M}^{-\frac{1}{3}} \Omega^{-1}\ .
\label{eq:cloudmass}
\end{equation}

These clouds move supersonically within the intercloud gas. If the intercloud gas
is ionized, typical Mach numbers are about 10. Therefore, we expect that the clouds
might be destroyed by Rayleigh-Taylor instabilities.
However, R\"{o}diger \& Hensler (2008) showed that these instabilities are suppressed
in the presence of (i) a sufficiently strong gravitational field of the clouds or (ii) a strong magnetic field.
The ram pressure exerted on the clouds
is $p_{\rm ram}=\rho_{\rm int} v_{\rm cl}^2$, where $\rho_{\rm int}$ is the intercloud
gas density and $v_{\rm cl}$ is the cloud velocity. 
Following  R\"{o}diger \& Hensler (2008) Rayleigh-Taylor instabilities are suppressed if
the gravitational acceleration $g$ is higher than the acceleration due to the drag by ram pressure
$a_{\rm D}$.
This yields approximately:
\begin{equation}
g \sim\frac{M_{\rm cl} G}{r_{\rm cl}^2} > a_{\rm D} \sim \frac{p_{\rm ram}}{\Sigma_{\rm cl}}\ .
\end{equation}
For the intercloud gas density we use the value of the Galactic Center $\rho_{\rm int}=10^{3}$~cm$^{-3}$
(Erickson et al. 1994).
Using $M_{\rm cl}=10$~M$_{\odot}$, $r_{\rm cl}=0.02$~pc, $v_{\rm cl}=100$~km\,s$^{-1}$,
the gravitational acceleration is about 4 times higher than the ram pressure drag.
If the intercloud gas has a 10 times higher density, only a magnetic field
with a field strength of $B \sim \sqrt{10\,\rho_{\rm int} v_{\rm cl}^2} \sim 1$~mG
can stabilize the clouds.
This kind of field strength is observed in the Circumnuclear Disk in the Galactic Center
(Plante et al. 1995). It is thus plausible that the torus clouds are stable
against Rayleigh-Taylor instabilities.

\section{Star formation in clumpy gas disks \label{sec:sfr}}

\subsection{Turbulent disks}

Following Paper II we assume that the star formation rate is proportional to the mean
density of the disk and the inverse of
the characteristic timescale for the cloud collapse, i.e.
the non-averaged local free fall time $t^{\rm cl}_{\rm ff}$:
\begin{equation}
\dot{\rho}_{*} \propto \frac{\rho}{t^{\rm cl}_{\rm ff}}\ .
\label{eq:rhostar}
\end{equation}
Since $t^{\rm cl}_{\rm ff} \propto \rho^{-\frac{1}{2}}$ 
this corresponds
to a Schmidt law of the form $\dot{\rho}_{*} \propto \rho^{\frac{3}{2}}$.
The factor of proportionality is given by the probability to find a
self-gravitating cloud, i.e. the volume filling factor $\phi_{\rm V}$.
Thus, the star formation rate is given by
\begin{equation}
\dot{\rho}_{*} = \phi_{\rm V} \frac{\rho}{t^{\rm cl}_{\rm ff}}=
\sqrt{\phi_{\rm V}} \frac{\rho}{t_{\rm ff}^{\rm H}}\ .
\end{equation}
Furthermore, we assume that stars are only born in the midplane of the disk
in regions that have the size of the turbulent driving length scale $l_{\rm driv}$,
because the clouds can collapse only within the turbulent
timescale $t_{\rm turb}=l_{\rm driv}/v_{\rm turb}$. We thus obtain
\begin{equation}
\dot{\Sigma}_{*}=\dot{\rho}_{*}\,l_{\rm driv}
\label{eq:starform}
\end{equation}
for the mass surface density turned into stars.

\subsection{Collisional disks}

For the collisional disk we assume that the star formation rate is
proportional to the overall density $\rho$ and the cloud collision
frequency $t^{-1}_{\rm coll}= \vartheta^{-1} \Omega$:
\begin{equation}
\dot{\rho}_{*} = \eta \rho \vartheta^{-1} \Omega\ ,
\end{equation}
where $\eta$ is an a priori unknown efficiency factor.
In terms of stellar mass per time this gives
\begin{equation}
\dot{M}_{*} = \eta M_{\rm gas} \vartheta^{-1} \Omega\ .
\end{equation}

\section{Thick disks in a generic galactic center}

The theoretical model described in the previous sections can now 
be applied to recent near-IR high-spatial resolution SINFONI observations
of nearby AGN.
Davies et al. (2007) showed that there had been recent
star formation in the central few tens of parsecs in a sample of nearby AGN.
Following on from this, Hicks et al. (2008) showed that distribution
and kinematics of the central concentrations of gas were similar to
those of the stars, and was geometrically thick with
$v_{\rm turb}/v_{\rm rot} > 1/2$.
They argued that this gas comprised the large scale
structure of the tori, implying that tori can form stars.
And they gave an estimate for the gas mass in this region as 10\% of
the dynamical mass.
Before we derive the physical disk parameters for the individual AGNs observed
by Davies et al. (2007), we give an overview over the different types of thick tori.
Note that $\vartheta$ is the disk transparency, i.e. $\vartheta \leq 1$ implies an opaque disk
whereas a large $\vartheta$ results in a transparent disk.
For a generic galactic center we assume a dynamical mass 
(which for radii $R$ larger 
than a few pc is dominated by the stellar content), 
of $M_{\rm dyn}=10^{8}$~M$_{\odot}$ and a gas mass of $M_{\rm gas}=10^{7}$~M$_{\odot}$, 
both within a radius of $R=20$~pc. The enclosed mass leads to 
a rotation velocity of $150$~km\,s$^{-1}$ at this radius. In addition, we adopt a cloud internal sound speed 
of $c_{\rm s}=1.5$~km\,s$^{-1}$.
The value for the sound speed corresponds to 
a cloud temperature of $\sim 500$~K when only thermal gas pressure is considered.
The sound speed is a measure of the pressure support against self-gravity and additional 
contributions to the pressure gradient inside clouds like magnetic fields may contribute.
The adopted sound speed leads to cloud masses of $M_{\rm cl} \propto c_{\rm s}^4 \sim 10$~M$_{\odot}$. 
The disk is stable with a Toomre parameter of $Q=4.7$ 
for a turbulent velocity supporting the vertical thickness of $v_{\rm turb} = 70$\,km/s.
Moreover, we assume a star formation rate of $\dot{M}=0.1$~M$_{\odot}$yr$^{-1}$.
This can be considered an upper limit to the current star
formation rate in the central few tens of parsecs. 
Davies et al. (2007) showed that while star formation had occured
there recently, it has now ceased. 
Based on the exponentially decaying starburst model they used and the
time averaged star formation rates they estimated, the current rates
are expected to be below this limit.

In the following we generate 
a generic set of models which (i) can describe
different evolutionary phases of a $\sim 20$~pc scale gas disk in terms
of mass accretion rate, disk thickness/transparency, and star formation rate
and (ii) can reproduce the observations at a generic time 
in the evolutionary sequence.
We first compare different disk models, representing evolutionary stages, to
the observations of 
the current state of the disk (Sec.~\ref{sec:massturb}--\ref{sec:ltcd}).
It is shown that a massive accretion event leads to a relatively thin turbulent
disk which forms stars at a high rate (SN model). Once the SN explode the
intercloud medium is blown out, leaving only dense, compact clouds. This results in a
collisional disk that may assume one of 
several states (Sec.~\ref{sec:mocd}--\ref{sec:ltcd})
depending on the external
mass accretion rate $\dot{M}$, the Toomre $Q$, and transparency $\vartheta$ parameters.
In Sec.~\ref{sec:tevol} we then investigate the evolution of the collisional disk.

\subsection{A massive turbulent disk creates a starburst \label{sec:massturb}}


We first try applying the turbulent SN disk model. With
the parameters described above it would have a mass accretion rate
\begin{equation}
\dot{M}=2.18\,\dot{M}_{*}^{-1} M_{\rm gas}^2 v_{\rm turb}^4 R^{-2} \xi^{-1} 
\end{equation}
in excess of $10^3\,$M$_{\odot}\,$yr$^{-1}$ and a ratio between the turbulent driving and 
dissipation length scales of
\begin{equation}
\delta=1.17\,\dot{M}_*^{-3}G^3M_{\rm gas}^6v_{\rm turb} R^{-6} \xi^{-2}> 10^4\ .
\end{equation}
Such a model can be discarded, 
because its mass accretion rate is far too high with respect to any supply from the outer galaxy and 
the lifetime $M_{\rm gas}/\dot{M} = 10^4$\, yr is shorter than the dynamical time $\Omega{-1} \sim 10^5$\,yr.

On the other hand, if we assume that a turbulent SN disk is responsible for the
starburst with a star formation rate of $\dot{M}_{*}=1$~M$_{\odot}$yr$^{-1}$ 
(comparable to the initial rate inferred by Davies et al. 2007) and that 
the turbulent 
velocity was lower ($v_{\rm turb}=20$~km\,s$^{-1}$) than the present collisional disk,
we find a mass accretion rate of $\dot{M}=2$~M$_{\odot}$yr$^{-1}$ and
$\delta=23$. These values are close to the values for the large-scale Galactic disk (see Paper II).
A typical cloud has a radius of $r_{\rm cl}=0.1$~pc and a mass of $M_{\rm cl}=10^4$~M$_{\odot}$.
This kind of disk contains about a thousand clouds within $R=20$~pc.

We conclude that a viable turbulent massive gas disk has a rather low turbulent velocity and is 
therefore moderately thin with $H/R = v_{\rm turb}/v_{\rm rot} \sim 0.13$. It then
yields large star formation rates of the order of one solar mass per year.
This represents a starburst which subsequently will destroy the disk once the 
supernovae explode after about 10~Myr. These explosions do not cause any harm to the
densest and most massive clouds, but they clear the space between the clouds, i.e.
they remove the initial intercloud medium. We are then left with a collisional disk.

\subsection{A massive, opaque, collisional disk\\ ($Q=5$, $\vartheta=1$) \label{sec:mocd}}

The collisional disks can be distinguish by their $Q$ and $\vartheta$ parameters. 
We  start with collisional and opaque disks, i.e. the mean free path of the
clouds is about the height of the disk $H$. This implies that along a vertical path through 
the disk there is on average one intervening cloud. Along a path in the disk midplane towards 
the center there are typically $N \sim 10$ clouds blocking the direct view. Collisions are 
frequent in such a torus or disk.
This yields a mass accretion rate of  
\begin{equation}
 \dot{M} = 2 \frac{v_{\rm turb}^3}{G Q \vartheta} = 3.4\;M_{\odot}{\,\rm yr}^{-1}\;.
\end{equation}
The volume filling factor of the clouds is $\Phi_{\rm V}=0.004$, the clouds have typical radii
of $r_{\rm cl}= 0.02$~pc, and masses of $M_{\rm cl}=15$~M$_{\odot}$. 

\subsection{A massive, transparent, collisional disk\\ ($Q=5$, $\vartheta=10$)}

The large mass accretion rate in the above model (Sec.~\ref{sec:mocd})
depends linearly on the collision rate and is a consequence of the low transparency.
If the disk is more transparent, $\vartheta=10$,  the mass accretion rate is therefore 
$\dot{M}=0.3$~M$_{\odot}$yr$^{-1}$, the volume filling factor of the clouds is $\Phi_{\rm V}=4\,10^{-5}$, and 
the clouds have typical radii
of $r_{\rm cl}= 0.002$~pc and masses of $M_{\rm cl}=1.5$~M$_{\odot}$.
We see that for the same Toomre-$Q$ a larger transparency implies smaller and 
less massive clouds with smaller volume filling factors.

\subsection{A light, opaque, collisional disk\\ ($Q=50$, $\vartheta=1$)}

If clouds are large and less dense we can have a disk which is light but still optically 
opaque due to dust in the clouds.
For this disk class we assume a gas mass of only $1\%$ of the dynamical mass, i.e. $M_{\rm gas}=10^6$~M$_{\odot}$.
This yields a mass accretion rate of $\dot{M}=0.3$~M$_{\odot}$yr$^{-1}$.
The volume filling factor is $\Phi_{\rm V}=0.04$, and the typical cloud radii and masses
are $r_{\rm cl}=0.2$~pc and $M_{\rm cl}=150$~M$_{\odot}$.

\subsection{A light, transparent, collisional disk\\ ($Q=50$, $\vartheta=10$) \label{sec:ltcd}}

The last disk class is transparent, $\vartheta=10$, and has a small gas mass, $M_{\rm gas}=10^6$~M$_{\odot}$.
It has the lowest mass accretion rate, $\dot{M}=0.03$~M$_{\odot}$yr$^{-1}$, and a low
volume filling factor, $\Phi_{\rm V}=4\,10^{-4}$. The cloud radii and masses are the same as
those of the massive, opaque disk ($r_{\rm cl}= 0.02$~pc and $M_{\rm cl}=15$~M$_{\odot}$).

We conclude that (i) a high mass accretion leads to a massive, opaque disk and (ii)
a ($Q=5$, $\vartheta=1$)-disk and a ($Q=50$, $\vartheta=10$)-disk share clouds
of the same mass and size. These clouds are very similar to those found in the Galactic Center
(see Vollmer et al. 2004). We thus can draw up  a disk evolution in which a massive, opaque
disk evolves with time into a light, transparent disk.

For the rest of the article we assume that the cloud mass of all disks is $M_{\rm cl}=10$~M$_{\odot}$.
This is equivalent to 
a common column density of all clouds 
of
\begin{equation}
N_{\rm cl}=\frac{3}{4} \frac{\Omega v_{\rm turb} \vartheta}{\pi G Q}\ .
\end{equation}
A thick disk with $\vartheta/Q=1/5$ has clouds with column densities of $N_{\rm cl} \sim 10^{24}$~cm$^{-2}$.

\section{Torus evolution \label{sec:tevol}}

The collisional disks described above can be identified with 
the observed $30$~pc-scale gas concentrations observed by Davies et al. (2007) and Hicks et al. (2008) 
which arguably correspond to the large scale structure of AGN tori.

It is now investigated how a collisional torus can evolve from a massive to a less massive state.
To do so we assume that the gas mass of the torus is proportional to its mass accretion rate
\begin{equation}
M_{\rm gas}=2^{-\frac{1}{3}} G^{-\frac{2}{3}} \vartheta^{\frac{1}{3}} Q^{-\frac{2}{3}} f^{-\frac{1}{3}} 
\dot{M}^{\frac{1}{3}} v_{\rm rot} R=\gamma \dot{M}^{x}\ .
\end{equation}
In this way we subsume all possible time dependencies of various parameters and their correlations in 
the time evolution of $\dot{M}$.
Together with Eq.~\ref{eq:cloudmass} this yields
\begin{equation}
\vartheta=\frac{\pi^4}{64} \frac{c_{\rm s}^4 R^2}{\gamma M_{\rm cl} G^2} \frac{1}{\dot{M}^x}\ .
\end{equation}
The expressions for the Toomre parameter, the turbulent velocity, and the star formation rate are then
\begin{equation}
Q=\frac{\pi^2}{\sqrt{128}} \frac{c_{\rm s}^2 R^4 \Omega^{\frac{3}{2}}}{\gamma^2 G^2 M_{\rm cl}^{\frac{1}{2}} f^{\frac{1}{2}}} \frac{1}{\dot{M}^{2x-\frac{1}{2}}}\ ,
\end{equation}
\begin{equation}
v_{\rm turb}=\frac{\pi^2}{\sqrt{128}} \frac{c_{\rm s}^2 R^2 \Omega^{\frac{1}{2}}}{\gamma M_{\rm cl}^{\frac{1}{2}} G f^{\frac{1}{2}}} \frac{1}{\dot{M}^{x-\frac{1}{2}}}\ ,
\label{eq:vturbev}
\end{equation}
and
\begin{equation}
\dot{M}_*=\eta \frac{64}{\pi^4} \frac{M_{\rm cl} \gamma^2 G^2 \Omega}{c_{\rm s}^4 R^2} \dot{M}^{2x}\ .
\end{equation}
In this way the behavior of the transparency, Toomre parameter, thickness of the disk via 
$H = v_{\rm turb}/\Omega$, and star formation rate can be identified with the change of the 
external mass accretion rate. For our stationary equilibrium disks to be applicable we must 
require the changes of the external mass accretion to be slow, so that the whole disk can 
adjust to the changing external conditions. This time for adjustment $t_{\rm eq}$ is approximately
the ratio between the torus gas mass and mass accretion rate. For typical values 
of $M_{\rm gas} \sim 10^7$~M$_{\odot}$ (Tab.~\ref{tab:input}) and 
$\dot{M} \sim 1$~M$_{\odot}$yr$^{-1}$ (Tab.~\ref{tab:prestorus}) this leads
to $t_{\rm eq} \sim 10$~Myr. This is smaller than the observed starburst ages (Tab.~\ref{tab:input}) 
for all AGNs except NGC~1097 where the time of adjustment and the starburst age are
comparable.

\subsection{Evolution at constant thickness}

Whenever the disk thickness stays the same during its evolution,  $v_{\rm turb}=$const. in time,
and subsequently $x=\frac{1}{2}$ (Eq.~\ref{eq:vturbev}),
this implies that the external mass accretion rate stays at a high level.
The gas mass, star formation rate, $\vartheta$, and the Toomre $Q$ parameter depend on the mass accretion rate
$\dot{M}$ in the following way:
\begin{equation}
M_{\rm gas} \propto \dot{M}^{\frac{1}{2}}\ ,\ \ \dot{M}_{*} \propto \dot{M}\ , \ \ \vartheta \propto \dot{M}^{-\frac{1}{2}}
\ , \ \ Q \propto \dot{M}^{-\frac{1}{2}}\ .
\end{equation}
The dependence of these parameters on the gas mass is then:
\begin{equation}
\dot{M}_{*} \propto M_{\rm gas}^2\ , \ \ \vartheta \propto M_{\rm gas}^{-1} \ , \ \ Q \propto M_{\rm gas}^{-1}
\label{eq:consthevol}
\end{equation}
and the relation for the cloud mass and the volume filling factor are:
\begin{equation}
M_{\rm cl} = {\rm const.}\ , \ \ \Phi_{\rm V} \propto M_{\rm gas}\ .
\end{equation}
It is remarkable that the star formation rate is linearly coupled to the mass accretion rate in 
this scenario.

\subsection{Evolution at constant mass}

Alternatively, at later stages when the mass accretion rate is low and changes little,
the torus mass may stay constant, $M_{\rm gas}=$const, and subsequently $x=0$ (Eq.~\ref{eq:vturbev}).
The turbulent velocity dispersion, star formation rate, $\vartheta$, 
and the Toomre depend on the mass accretion rate
$\dot{M}$ now in the following way:
\begin{equation}
v_{\rm turb} \propto \dot{M}^{\frac{1}{2}}\ , \ \ \dot{M}_{*}={\rm const.}\ , \ \ 
\vartheta={\rm const.}\ , \ \ Q \propto \dot{M}^{\frac{1}{2}}\ .
\label{eq:constmevol}
\end{equation}
The relation for the cloud mass and the volume filling factor are:
\begin{equation}
M_{\rm cl} = {\rm const.}\ , \ \ \Phi_{\rm V} \propto \dot{M}^{-\frac{1}{2}}\ .
\end{equation}
At constant disk mass and decreasing external mass supply the disk will become geometrically 
thin without changing the transparency. 

An interesting result is that during both types of torus evolution---at constant turbulent
velocity  and at constant gas mass---the cloud mass does not change.

\section{Torus evolution scenarios \label{sec:scenarios}}

In the picture of quasi-stationary equilibrium disks driven by cloud collisions the evolution will be 
determined by the external mass accretion rate,
i.e. the mass inflow from distances $>100$~pc. 
We divide the torus evolution into three phases:
\begin{itemize}
\item
{\it Phase I: Initial massive infall and formation of a turbulent, massive gas disk}:\\
An initial rapid infall of a large amount of gas , $M_{\rm gas} \sim 10^{7}$~M$_{\odot}$,
within a short time ($\Delta t < 1$~Myr; $\dot{M} > 10$~M$_{\odot}$yr$^{-1}$)
leads to the formation of a massive ($Q \sim 1$), moderately thin ($v_{\rm turb}/v_{\rm rot} < 5$)
gas disk in which star formation proceeds. This phase will be recognized as a starburst.
The disk becomes turbulent and the turbulence is maintained by the energy input from
stellar winds. After $\sim 10$~Myr the first SN explode and will rapidly remove the initial 
intercloud medium from the
disk. Only the most massive and densest clouds which are not Jeans unstable will survive.
This leads in the following to a {\it collisional torus}.
\item
{\it Phase II: Torus evolution at constant turbulent velocity}:\\
During the first phase of its evolution the massive collisional torus stays thick. This implies that the
mass accretion rate within the torus, and thus also the external mass accretion rate $\dot{M}$, 
do not decrease significantly during this phase. The gas mass and
the cloud collision rate ($t_{\rm coll} \propto \vartheta^{-1}$) decrease with decreasing $\dot{M}$, 
whereas $Q$ increases with the square root of the external mass accretion rate.
The star formation rate within the torus decreases with decreasing $\dot{M}$ (Eq.~\ref{eq:consthevol}).
\item
{\it Phase III: Torus evolution at constant gas mass}:\\
Once the external mass accretion rate has significantly decreased, the torus
evolves at constant gas mass. The velocity dispersion and $Q$ decrease  
with the square root of the external mass accretion rate, whereas the
cloud collision rate ($t_{\rm coll} \propto \vartheta$) and the star formation rate stays constant
(Eq.~\ref{eq:constmevol}). The Circumnuclear Disk (CND) in the Galactic Center
(G\"{u}sten et al. 1987) represents such a late stage ($Q=190$, $\vartheta=15$; Vollmer et al. 2004) 
of the torus to disk evolution.
\end{itemize}

\begin{figure}
        \resizebox{\hsize}{!}{\includegraphics{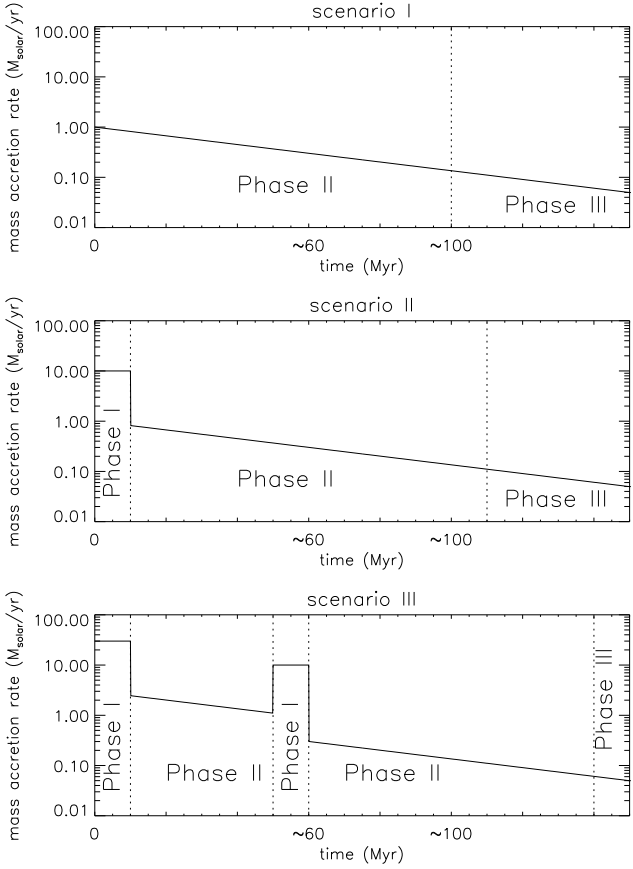}}
        \caption{Schematic of torus evolution scenarios: Torus mass accretion rate (M$_{\odot}$yr$^{-1}$) is plotted 
	  as a function of time (Myr). Phase I:  A massive ($Q \sim 1$) turbulent disk is associated
	  with a starburst. Phase II: A collisional torus evolves at a constant turbulent
	  velocity dispersion, i.e. thick torus evolution. Phase III: The torus becomes thin
	   and evolves at an approximately constant gas mass.
        } \label{fig:scenarios}
\end{figure}

Depending on the time evolution of the external mass accretion rate (from distances
larger than $100$~pc) we suggest three possible evolution scenarios (Fig.~\ref{fig:scenarios}):
\begin{itemize}
\item
{\it Scenario I:}\\
The torus never reaches Phase I, because (i) it is already 
clumpy from 
the very beginning,
(ii) there is already star formation occurring at scales of $\sim 10$~pc from the galactic center;
the associated SN explosions and/or winds remove the existing intercloud medium and/or
inhibit the formation of an intercloud medium. The two possibilities imply that the
initial mass accretion rate is not very high. A third possibility (iii) is that the disk's velocity
dispersion is prohibitively 
large to allow 
a $Q=1$ disk. In this case star formation proceeds via
cloud-cloud collisions. The star formation rate is lower than that of a massive turbulent disk.
The rate of momentum injection due to SNe and stellar winds is
\begin{equation}
F=p_{*} \Delta A=5\,10^{33} (\frac{\dot{M}_{*}}{{\rm M_{\odot}yr^{-1}}})\ {\rm dyne}\ ,
\end{equation}
where $p_{*}$ is the pressure due to SNe and stellar winds exerted on a surface $\Delta A$
(Veilleux et al. 2005). Assuming the same initial turbulent velocity for the clouds and 
the intercloud medium, the intercloud medium can be removed from the disk if
$p_{*} \geq \rho_{\rm IM} \Phi_{\rm V}^{\rm IM} v_{\rm turb}^2$, where $\rho_{\rm IM}$ and $\Phi_{\rm V}^{\rm IM}$are the density and the volume filling factor of the intercloud medium.
Assuming $\dot{M}_{*}=0.02$~M$_{\odot}$yr$^{-1}$, $R=10$~pc (Sect.~\ref{sec:formeff}), and
$v_{\rm turb}=50$~km\,s$^{-1}$ (Tab.~\ref{tab:input}) leads to 
$n_{IM} \Phi_{\rm V}^{\rm IM} \leq 10^3$~cm$^{-3}$.
Since the disk volume averaged density of the clouds is $n_{cl} \Phi_{\rm V}=10^4$~cm$^{-3}$
(Tab.~\ref{tab:prestorus}), the intercloud space can only be cleared by
SN explosions and stellar winds if the intercloud medium contains less than $\sim 10$\,\% of the
total disk mass.
\item
{\it Scenario II:}\\
Due to an initial, massive infall a massive ($Q \sim 1$) turbulent star-forming disk is formed
(Phase I). The turbulence in this disk is maintained through the energy supply by 
feedback from rapid star formation. 
The subsequent SN explosions destroy the disk structure after $10$~Myr,
i.e. the intercloud medium is removed
leaving only the densest, most massive clouds which remain Jeans-stable. The disk becomes collisional
and stays geometrically thick (Phase II). After $\sim 100$~Myr the mass accretion rate decreases and the
disk becomes thin (Phase III) and ultimately transparent
($\vartheta > 5$). The time at which the torus changes from Phase II into Phase III 
depends on the time evolution of the external mass accretion rate in this scenario.
\item
{\it Scenario III:}\\
Due to an initial massive infall event, a massive ($Q \sim 1$) turbulent star-forming disk appears
(Phase I). As in Scenario II,  SN explosions destroy the disk
structure after $\sim 10$~Myr, the intercloud medium is removed 
 and only the densest, most massive clouds are left over, which are Jeans-stable. The disk becomes collisional
and will stay thick as long as  the external mass accretion rate is sufficiently high (Phase II).
Due to a secondary massive and rapid gas infall a second massive ($Q \sim 1$) turbulent star-forming disk 
can form (Phase I), which evolves again into a collisional disk after $\sim 10$~Myr (Phase II).
After $\sim 100$~Myr the mass accretion rate has sufficiently decreased and the 
disk becomes thin (Phase III) and ultimately transparent
($\vartheta > 5$). The time at which the torus changes from Phase II into Phase III 
depends again on the time evolution of the external mass accretion rate.
\end{itemize}

\section{Applying the observations \label{sec:comparison}}

The scenarios of Sect.~\ref{sec:scenarios} derived from our analytical models can now be compared 
with the VLT SINFONI observations of Davies et al. (2007) 
and Hicks et al. (2008).
As has been shown in Sect.~\ref{sec:massturb} the present state of the disk
can only be described consistently by a collisional disk model.
The comparison with observations will allow us to derive the parameters of (i) the
present torus, (ii) the initial torus immediately after Phase I, (iii) 
the massive ($Q \sim 1$) turbulent gas disk that gave rise to the initial starburst (scenario II and III),
and (iii) the star formation efficiency of the collisional phase.

The observables, i.e. the input parameters for our analytical model, are the
radius from the galactic center $R$, the rotation velocity $v_{\rm rot}$, the turbulent velocity
dispersion $v_{\rm turb}$, the gas mass $M_{\rm gas}$, the peak star formation rate during the initial starburst $\dot{M}_{*}^{\rm peak}$, and
the age of the initial starburst $ t_{\rm SB}$. We assume for all AGN, except NGC~1097, a starburst duration of $10$~Myr,
i.e. the the starburst continues until the first SN explode. 
We only apply our model to a subsample of 6 nearby AGNs from Davies et al. (2007) for which
these input parameters are sufficiently well known.
The parameters for these objects can be found in Table~\ref{tab:input}.
For each object, the table gives its Seyfert type and the radius
within which the subsequent parameters apply.
The dispersion $v_{\rm turb}$ is the value measured from the data (after
accounting for instrumental broadening), while the rotation velocity
$v_{\rm rot}$ is a Keplerian equivalent value.
This means that it represents the rotation velocity that would be
needed if ordered circular motions in a single plane supported the entire
dynamical mass. 
It is therefore significantly greater than the measured rotation speed.
The gas mass $M_{gas}$ is difficult to derive.
Hicks et al. (2008) estimated it from a combination of diagnostics,
including the 1.3\,mm CO luminosity, the 2.12\,$\mu$m H$_2$ 1-0\,S(1) 
luminosity, and a
comparison to the gas-to-dynamical mass ratios in other spiral and
starburst galaxies.
The peak star formation rates $\dot{M}_{*}^{\rm peak}$ are derived
from the starburst models used in Davies et al. (2007).
They are simply the star formation rates required to form the young
stars in a timescale of 10\,Myr (in contrast to the time-averaged rates
given in that paper).
The only exception is NGC\,3783, for which we have adopted a
2-starburst model with ages of 110\,Myr and 30\,Myr.
Such models were not considered by Davies et al. (2007) because of the
limited number of diagnostics.
In our scenario with stronger theoretical constraints on the models we need 
these additional degrees of freedom to reproduce the observations.
The final column gives the age $t_{\rm SB}$ of the most recent
starburst.
All equations to derive the properties of the present collisional torus,
the massive $(Q \sim 1)$ turbulent gas disk, and the initial collisional torus
are given in Table~\ref{tab:equations}.


\begin{table*}
      \caption{Input parameters from Davies et al. (2007) and Hicks et
      al. (2008); see
      Sect.~\ref{sec:comparison} for details}
         \label{tab:input}
      \[
       \begin{array}{|l|l|c|c|c|c|c|c|}
        \hline
        {\rm name} & {\rm type} & R  & v_{\rm rot} & v_{\rm turb} & M_{\rm gas} & \dot{M}_{*}^{\rm peak} & t_{\rm SB} \\
	 & & {\rm (pc)} & {\rm (km\,s^{-1})} &  {\rm (km\,s^{-1})} & (10^6\ {\rm M}_{\odot}) & ({\rm M}_{\odot}{\rm yr}^{-1}) & ({\rm Myr}) \\
	\hline
	\hline
	{\rm Circinus} & {\rm Sy2} & 10 & 93 & 56 & 2 & 0.02 & 80 \\
 	\hline
	{\rm NGC~3783} & {\rm Sy1} & 30 & 88 & 35 & 5 & 0.3 & 30 \\
	\hline
	{\rm NGC~3227} & {\rm Sy1} & 30 & 173 & 95 & 20 & 5.0 & 40 \\
	\hline
	{\rm NGC~1068} & {\rm Sy2} & 30 & 184 & 88 & 20 & 14.0 & 200 \\
	\hline
	{\rm NGC~1097} & {\rm Sy1} & 30 & 122 & 59 & 10 & 1.0 & 8 \\
	\hline
	{\rm NGC~7469} & {\rm Sy1} & 30 & 123 & 63 & 10 & 3.0 & 110 \\
	\hline
        \end{array}
      \]
\end{table*}

\begin{table}
      \caption{Equations to derive the disk/torus properties from observations. 
	Velocities are in units of km\,s$^{-1}$, radii in pc,
	masses in M$_{\odot}$, and star formation rates in M$_{\odot}$yr$^{-1}$.}
         \label{tab:equations}
      \[
       \begin{array}{|l|}
        \hline
        {\rm present\ collisional\ torus}\\
	\hline
	 \\
	Q^{\rm pr}=(v_{\rm turb}^{\rm pr}/v_{\rm rot})\,(M_{\rm dyn}/M_{\rm gas}^{\rm pr})\\
	 \\
	\vartheta^{\rm pr}=1.9 \times 10^{-9} Q^{\rm pr} R/(v_{\rm rot} v_{\rm turb}^{\rm pr} M_{\rm cl})\\
	 \\
	\dot{M}^{\rm pr}=2\,(v_{\rm turb}^{\rm pr})^{3}/(G Q^{\rm pr} \vartheta^{\rm pr})\\
	 \\
	\hline
	{\rm massive\ turbulent\ gas\ disk}\\
	\hline
	 \\
	Q^{\rm disk}=9674\,\dot{M}_*^{-\frac{2}{11}} v_{\rm rot}\\
	 \\
	M_{\rm gas}^{\rm disk}=3.97 \times 10^{5} \dot{M}_*^{\frac{5}{11}} R\\
	 \\
	v_{\rm turb}^{\rm disk}=18.2\,\dot{M}_*^{\frac{3}{11}}\\
	 \\
	\hline
	{\rm initial\ collisional\ torus}\\
	\hline
	 \\
	\dot{M}^{\rm init}=\dot{M} (M_{\rm gas}^{\rm disk}/M_{\rm gas}^{\rm pr})^{2}\\
	 \\
	Q^{\rm init}=Q^{\rm pr} M_{\rm gas}^{\rm pr}/M_{\rm gas}^{\rm disk}\\
	 \\
	\vartheta^{\rm init}=\vartheta^{\rm pr}\,M_{\rm gas}^{\rm pr}/M_{\rm gas}^{\rm disk}\\
	 \\
	\hline
        \end{array}
      \]
\end{table}

\subsection{The present collisional torus}

We recall the assumption that all gas clouds have 
a constant mass of $M_{\rm cl}=10$~M$_{\odot}$.
Moreover 
the sound speed within the clouds is set to $c_{\rm s}=1.5$~km\,s$^{-1}$.
The Toomre $Q$ parameter of the present disks is directly calculated from Eq.~\ref{eq:tq}.
We then use the following expressions for $\vartheta$ and the mass accretion rate $\dot{M}$ 
which follow from the expression derived for the turbulent velocity dispersion in Paper I
$v_{\rm turb}=(\frac{1}{2} G \vartheta Q f^{-1} \dot{M})^{\frac{1}{3}}$ and Eq.~\ref{eq:cloudmass}
\begin{equation}
\vartheta=\frac{\pi^4}{64} c_{\rm s}^4 G^{-1} Q \Omega^{-1} M_{\rm cl}^{-1} v_{\rm turb}^{-1}\ ,
\label{eq:taupres}
\end{equation}
and
\begin{equation}
\dot{M}=2 f v_{\rm turb}^3 G^{-1} Q^{-1} \vartheta^{-1}\ .
\end{equation}
The torus mass accretion rates (Table~\ref{tab:prestorus}) 
include the geometric $f=0.1$ factor introduced in Sec.~\ref{sec:colldisk}. 
The resulting rates are still higher than the
black hole mass accretion rates and the wind mass loss rates discussed in Sec.~\ref{sec:fuel}). 
The parameters derived in this way for our sample of 6 AGNs are shown in Table~\ref{tab:prestorus}.

All tori show Toomre $Q$ parameters between 4 and 6, i.e. they are massive thick tori.
Moreover, 3 AGNs (Circinus, NGC~3227, and NGC~1068) have opaque tori ($\vartheta < 2$),
2 AGNs (NGC~1097 and NGC~7469) show moderately transparent tori ($\vartheta \sim 3-4$),
and 1 AGN (NGC~3783) has a transparent torus ($\vartheta > 5$).
It is worth noting that two out of the 3 AGNs with opaque tori are classified as Sy2.
We find the smallest mass accretion rate for NGC~3783. The highest mass accretion
rates (NGC~3227 and NGC~1068) are several ten times higher than that of NGC~3783.
\begin{table}
      \caption{Parameters of the present collisional tori.}
         \label{tab:prestorus}
      \[
       \begin{array}{|l|c|c|c|c|c|c|}
        \hline
        {\rm name} & \dot{M} & Q & \vartheta & \Phi_{\rm V} & \rho_{\rm cl} & r_{\rm cl}  \\
	 & ({\rm M}_{\odot}{\rm yr}^{-1}) & & & 10^{-3} & 10^7~{\rm (cm}^{-3}) & {\rm (pc)} \\
	\hline
	\hline
	{\rm Circinus} & 0.75 & 5.7 & 1.9 & 1.2 & 3.9 & 0.01 \\
 	\hline
	{\rm NGC~3783} & 0.07 & 4.1 & 6.8 &0.2 & 2.7 & 0.01 \\
	\hline
	{\rm NGC~3227} & 4.27 & 5.5 & 1.7& 0.5& 3.9 & 0.01 \\
	\hline
	{\rm NGC~1068} & 3.45 & 5.4 & 1.7 &1.2& 1.8 & 0.02\\
	\hline
	{\rm NGC~1097} & 0.59 & 4.8 & 3.4 &0.4& 2.6 & 0.02\\
	\hline
	{\rm NGC~7469} & 0.66 & 5.1 & 3.4 &0.6& 1.8 & 0.02\\
	\hline
        \end{array}
      \]
\end{table}

\subsection{The massive ($Q \sim 1$) turbulent gas disk}

We now derive the parameters of the initial massive turbulent gas disk which gave
rise to the initial starburst.
For this we apply the SN model (see Sect.~\ref{sec:theory}) where
the energy source for maintaining turbulence in the disk are
stellar winds. Their energy input is comparable to that of SN explosions 
(MacLow \& Klessen 2004). Therefore we do not need to change the
formalism of the SN model.
We further assume that the mass accretion rate equals the peak star formation rate
given by Davies et al. (2007) (Table~\ref{tab:input}) and
$\delta=5$, which is the Galactic value (Vollmer \& Beckert 2003).

We assume that for all AGNs, except for Circinus and NGC~3783, scenario II is valid.
For Circinus we argue below that scenario I applies, because its peak star formation rate is
a factor of more than 
10 lower when compared to 
the other starbursts.
This does not imply 
that star formation is not occurring, just that there was no initial
massive accretion event.
For NGC~3783 scenario III is more applicable, 
because a single starburst
leads to an enormously high initial mass accretion rate compared to the present value,
which we think is implausible.
The double starburst we have adopted for NGC~3783, which is consistent with the observations of
Davies et al. (2007), requires a first intense burst with
$\dot{M}_{*}=4.7$~M$_{\odot}$yr$^{-1}$ to have occurred 110\,Myr ago, followed by a
second distinct burst with $\dot{M}_{*}=0.3$~M$_{\odot}$yr$^{-1}$ only
30\,Myr ago.


The Toomre $Q$ parameter, the total gas mass, and the turbulent 
velocity dispersion can then be calculated using the following expressions:
\begin{equation}
Q=0.81\, G^{-\frac{2}{11}} \dot{M}_{*}^{-\frac{2}{11}} \delta^{-\frac{3}{11}} \xi^{-\frac{5}{22}} v_{\rm rot}\ ,
\end{equation}
\begin{equation}
M_{\rm gas}=8.5\,10^{-5} G^{-\frac{8}{11}} \dot{M}_{*}^{\frac{5}{11}} \delta^{-\frac{1}{11}} \xi^{\frac{1}{11}} R\ ,
\label{eq:mgasmassif}
\end{equation}
and
\begin{equation}
v_{\rm turb}=0.82\, G^{\frac{3}{11}} \dot{M}_{*}^{\frac{3}{11}} \delta^{-\frac{1}{11}} \xi^{\frac{1}{11}}\ ,
\end{equation}
where $\xi=4.6\,10^{-8}$~(pc/yr)$^2$ (Vollmer \& Beckert 2003).

Using a peak star formation rate of $\dot{M}_{*}=0.02$~M$_{\odot}$yr$^{-1}$ for Circinus 
leads to a gas mass of the massive turbulent disk which is smaller than that of the present
collisional torus. We therefore conclude that scenario II (Fig.~\ref{fig:scenarios}) does not apply
to this AGN. Instead, scenario I yields more appropriate results. 
The nuclear disk in Circinus did not experience a turbulent, supernovae and stellar wind driven $Q\sim1$ disk.

The gas masses of the turbulent starburst disks is between $10\%$ and $100\%$ 
higher than the gas mass of the corresponding present collisional torus, 
except for Circinus. In the course of torus evolution
(Fig.~\ref{fig:scenarios}) the loss of gas mass in the disk is moderate (up to a factor of 2;
Tab.~\ref{tab:input} and \ref{tab:turbdisk}).
The star forming disks are moderately thin ($v_{\rm rot}/v_{\rm turb} \sim R/H \sim 5$-$7$) and their
Toomre $Q$ parameter is close to unity.
\begin{table}
      \caption{Parameters of the initial massive ($Q \sim 1$) turbulent gas disk
	that gave rise to the initial starburst.}
         \label{tab:turbdisk}
      \[
       \begin{array}{|l|c|c|c|c|}
        \hline
        {\rm name} & M_{\rm gas} & \dot{M} & v_{\rm turb} & Q  \\
	  & (10^6\ {\rm M}_{\odot})   & ({\rm M}_{\odot}{\rm yr}^{-1}) & {\rm (km\,s^{-1})}&  \\
	\hline
	\hline
	{\rm Circinus} & 0.7 & 0.02 & 6 & 1.9 \\
 	\hline
	{\rm NGC~3783} & 6.9 & 0.3 & 13 & 1.1 \\
	\hline
	{\rm NGC~3227} & 24.8 & 5.0 & 28 & 1.3 \\
	\hline
	{\rm NGC~1068} & 40.0 & 14.0 & 37 & 1.2 \\
	\hline
	{\rm NGC~1097} & 11.9 & 1.0 & 18 & 1.2 \\
	\hline
	{\rm NGC~7469} & 19.6 & 3.0 & 25 & 1.0 \\
	\hline
        \end{array}
      \]
\end{table}

\subsection{The initial collisional torus}

Since all observed tori are thick (Table~\ref{tab:input}), they
are all in Phase II of their evolution (see Sect.~\ref{sec:scenarios}), i.e.
they evolve at constant thickness or velocity dispersion.
This implies the following relation between the gas mass and
the mass accretion rate:
\begin{equation}
\dot{M} \propto M_{\rm gas}^2\ .
\end{equation}
We can safely assume that the gas mass of  the initial collisional torus
at the end of Phase I is close to the gas mass of the massive
($Q \sim 1$) turbulent disk (Table~\ref{tab:turbdisk}).
For Circinus we estimate the gas mass of the initial collisional
torus by postulating that the disk was initially opaque, i.e. $\vartheta \sim 1$.
This leads to an initial gas mass of $M_{\rm gas}=3.4\,10^6$~M$_{\odot}$
at the beginning of Phase II.

For all other AGN a continuous transition from phase I to II provides
the mass accretion rate of the initial
collisional torus. The $Q$ and $\vartheta$ parameters can then be
calculated using the following expressions (see Sect.~\ref{sec:tevol}):
\begin{equation}
Q \propto M_{\rm gas}^{-1}\ , \ \ \ \vartheta \propto M_{\rm gas}^{-1}\ .
\end{equation}

Four of the initial collisional disks at the beginning of Phase II 
(Circinus, NGC~3227, NGC~1068, and NGC~7469) had
Toomre $Q$ parameters around $Q=3$ and $\vartheta$ close to unity,
i.e. they were massive and opaque. Two initial collisional disks
(NGC~3783 and NGC~1097) had been massive and transparent.
\begin{table}
      \caption{Parameters of the initial collisional tori.}
         \label{tab:initorus}
      \[
       \begin{array}{|l|c|c|c|}
        \hline
        {\rm name} & \dot{M} & Q & \vartheta \\
	 & ({\rm M}_{\odot}{\rm yr}^{-1}) & &  \\
	\hline
	\hline
	{\rm Circinus} & 2.1 & 3.4 & 1.1 \\
 	\hline
	{\rm NGC~3783} & 0.1 & 3.0 & 5.0 \\
	\hline
	{\rm NGC~3227} & 6.5 & 4.4 & 1.4 \\
	\hline
	{\rm NGC~1068} & 13.5 & 2.7 & 0.9 \\
	\hline
	{\rm NGC~1097} & 0.8 & 4.0 & 2.9 \\
	\hline
	{\rm NGC~7469} & 2.6 & 2.6 & 1.7 \\
	\hline
        \end{array}
      \]
\end{table}

\subsection{The evolution of the torus mass accretion rate \label{sec:tmassacc}}

\begin{figure}
        \resizebox{\hsize}{!}{\includegraphics{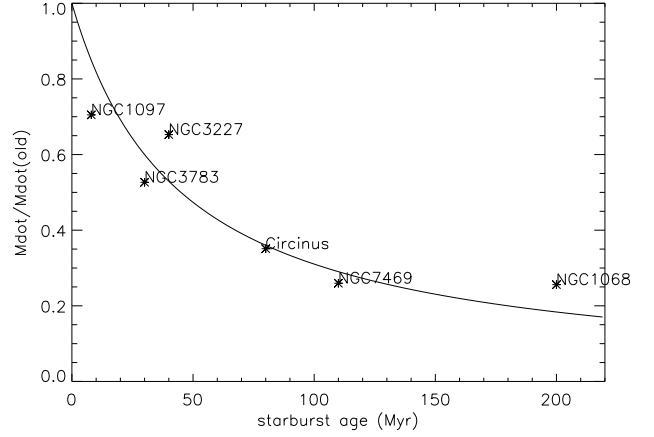}}
        \caption{Ratio between the mass accretion 
	  rate of the present collisional torus and that of the initial collisional torus 
	  (end of Phase I and beginning of Phase II) as a function of time.
	  The zero point in time is the end of Phase I ($\vartheta =1$ for Circinus).
        } \label{fig:tmassacc}
\end{figure}

In this section we investigate the evolution of the torus mass accretion rate
with time in Phase II (see Sect.~\ref{sec:scenarios}). 
For this we plot in Fig.~\ref{fig:tmassacc} the fraction between the mass accretion 
rate of the present collisional torus and that of the initial collisional torus 
(end of Phase I and beginning of Phase II) as a function of time for our sample.
To determine the time that a given torus passed in Phase II, we place ourselves in scenario II 
(Fig.~\ref{fig:scenarios}) and adopt the starburst ages of Davies et al. (2007) for all AGNs 
except NGC~3783. We assume scenario III for this galaxy, i.e. the occurrence of two
distinct starbursts. This leads to an estimated age of the most recent starburst of 30~Myr.
The solid line is meant to guide the eye. 
Based on  this plot we conclude that the
mass accretion rate shows a slow monotonic decrease with time.
It decreases to one fourth of its initial value in about 150~Myr.

\subsection{The star formation efficiency of collisional tori \label{sec:formeff}}

Star formation in collisional disks is expected (see Sec.~\ref{sec:sfr}) to be proportional to the
cloud collision rate $t_{\rm coll}^{-1}=\Omega/\vartheta$:
\begin{equation}
\dot{M}_{*}=\eta M_{\rm gas} \vartheta^{-1} \Omega\ ,
\end{equation}
where $\eta$ is the star formation efficiency which we would like
to determine. This is only possible for Circinus, because only in this
case the peak star formation rate derived from observations reflects the
initial star formation rate of the collisional torus (scenario II).
For all other AGNs the peak star formation rate is related to a 
massive ($Q \sim 1$) turbulent gas disk.
During the torus evolution at constant thickness (Phase II) the star formation
rate is proportional to the square of the gas mass 
\begin{equation}
\dot{M}_* \propto M_{\rm gas}^2\ .
\end{equation}
The initial and present gas mass of the torus in Circinus are $M_{\rm gas}^{\rm init}=3.4\,10^6$~M$_{\odot}$ and
$M_{\rm gas}=2\,10^6$~M$_{\odot}$; the peak star formation rate is $\dot{M}_*^{\rm init}=0.02$~M$_{\odot}$yr$^{-1}$.
This leads to an estimate of the present star formation rate of $\dot{M}_*=7\,10^{-3}$~M$_{\odot}$yr$^{-1}$.
The resulting star formation efficiency is 
\begin{equation}
\eta=\frac{R\dot{M}_*^{\rm init}}{M_{\rm gas}^{\rm init} v_{\rm rot}} \simeq 10^{-3}\ .
\end{equation}
For a turbulent galactic disk the star formation law is
\begin{equation}
\dot{M}_*=\eta_{\rm gal} M_{\rm gas} \Omega\ .
\end{equation} 
Thus, one has to compare $\eta/\vartheta \sim 5\,10^{-4}$ for a collisional disk with $\eta_{\rm gal}=0.017$ 
derived for galactic SN driven turbulent gas disk.
We conclude that the star formation in a collisional torus is about 10 times
less efficient than that of a turbulent gas disk of the same mass.
In our sample of 6 AGN the star formation rate of the present collisional torus is small compared
to the mass accretion rate. The average fraction between the mass accretion rate and
the star formation rate in these tori is $25 \pm 7$.

\subsection{Obscuring the starburst \label{sec:obscuring}}

\begin{figure}
        \resizebox{\hsize}{!}{\includegraphics{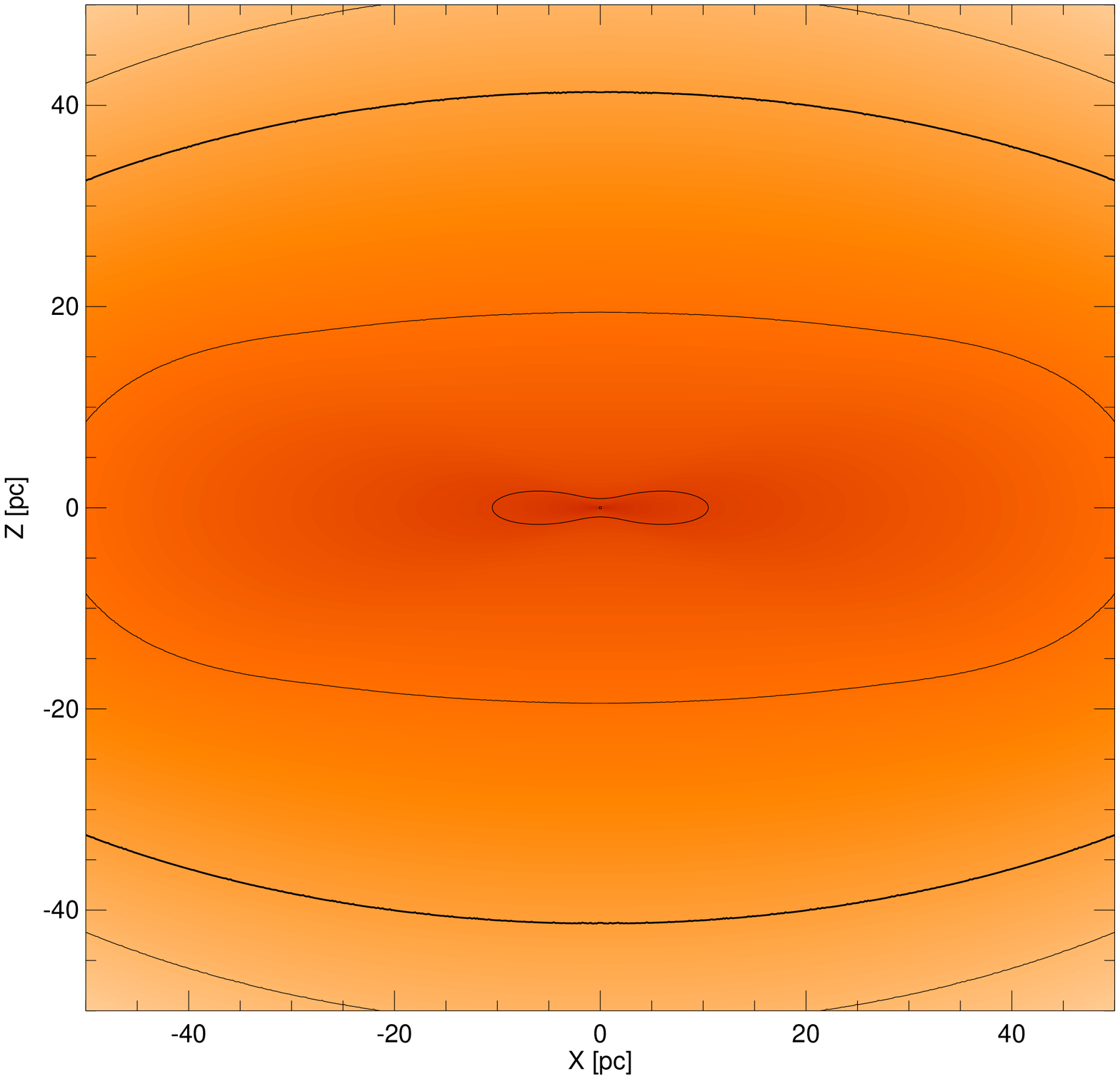}}
	\resizebox{\hsize}{!}{\includegraphics{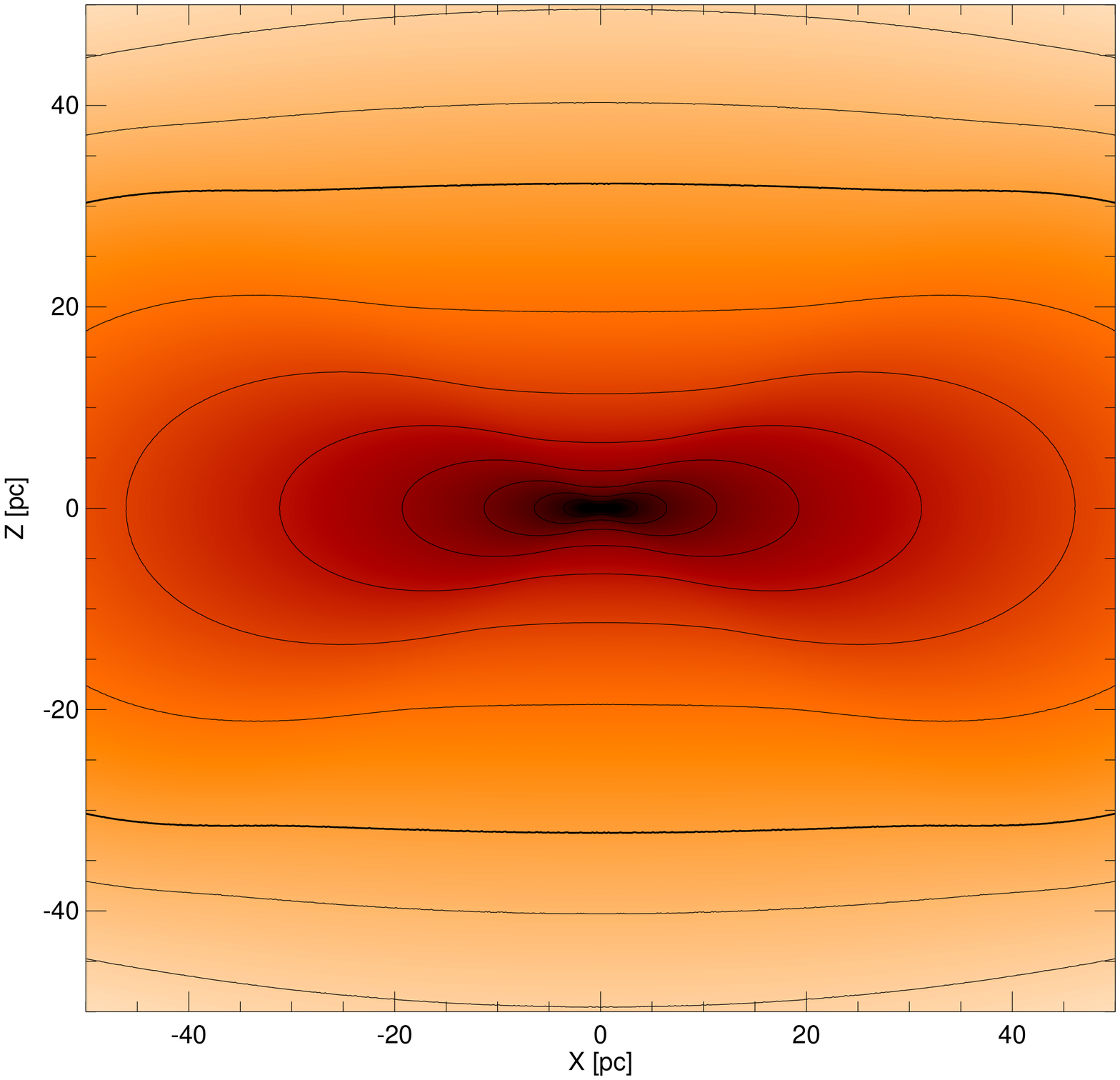}}
        \caption{Projected mean number of clouds $\Lambda$ along the line of sight
	  through the torus for the screen model. 
	  Contour levels are (0.5,0.75,1,1.5,2,2.5,3,..).
	  The thick contour is for $\Lambda=1$ corresponding to an non obscuration probability of $e^{-1}$.
	  Upper panel: torus inclination $i=45^{\circ}$. Lower panel:
	  torus inclination $i=60^{\circ}$ ($i$ measured from the torus axis). The size of the box is $100$~pc.
        } \label{fig:Lambda}
\end{figure}

The stellar luminosities calculated by Davies et al. (2007; their
Fig. 8) indicate that on scales of several tens of parsecs, the
starbursts can only be weakly obscured.
The reason for this is explained by Davies et al. (2006), and
summarised here.
If one assumes no obscuration, the starbust already comprises
typically a few percent of the galaxy's bolometric luminosity.
Even a moderate optical depth will lead to nearly all the optical
light being absorbed and re-radiated in the far-infrared; and in
addition the scaling one derives for the starburst would increase.
The limiting constraint is that the starburst luminosity cannot exceed the galaxy's bolometric
luminosity.
We estimate the resulting extinction (for screen models) to be in the
range $A_{\rm V}=4$ to $8$.
This is consistent with a clumpy torus with area filling factors below unity (transparency $\vartheta > 1$) for
radii larger than about $1$~pc. The optical extinction is then mainly
due to dust located in the foreground of the star forming region.
In Sy2 galaxies, however, not only the central AGN is obscured by the torus, but also 
the inner part of the torus and embedded star forming regions are self-obscured by the extended torus.
To verify this idea, we use the model of Beckert \& Duschl (2004) and H\"{o}nig et al. (2006) which is
based on the formalism developed by Vollmer et al. (2004) to estimate the extinction
using (i) a screen model where the starburst occurred in the torus mid-plane behind half the
torus, and (ii) a mixed model where the forming stars and the clouds share the same spatial distribution 
within the torus.
As an example we use parameters most appropriate for  NGC~3227 or NGC~1068: 
$v_{\rm turb}/v_{\rm rot} = 0.55$, $\vartheta = 2$ with an outer radius of 80~pc. 
Fig.~\ref{fig:Lambda} shows the mean number of clouds $\Lambda$ along the line of sight
through the torus for two torus inclinations in the case of a screen model.
\begin{equation}
\Lambda=\int l_{\rm coll}^{-1} {\rm d}s\ ,
\end{equation}
where $l_{\rm coll}=\vartheta\,H$ is the mean free path of the clouds and $H$ the disk height.
Obscuration with $\Lambda \ge 3$ leading to an likelihood of non-obscuration of less than $ e^{-3}$ occurs 
at a radii $R \le 20$~pc for $i=60^{\circ}$ and not at all for $i=40^{\circ}$. 
For lower inclinations the obscuration by clouds is ineffective. For $i >60^{\circ}$ the obscuration 
pattern does not change much for geometrically thick tori. 

\begin{figure}
        \resizebox{\hsize}{!}{\includegraphics{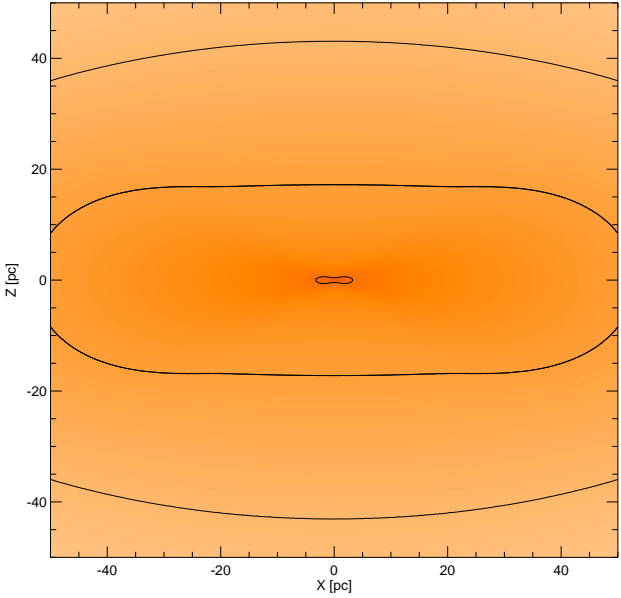}}
	\caption{Ratio between the extinction-free starburst emission to the emission
	  with cloud absorption $I_0/I_{\rm ext}$ for an torus inclination of
	    $i=50^{\circ}$. Contour levels are (1.5,2,2.5). The thick contour
	    is for $I_0/I_{\rm ext}=2$. It is the seconds outermost contour.
	  } \label{fig:mix}
\end{figure}

Fig.~\ref{fig:mix} shows the ratio between the extinction-free starburst emission and the emission
with obscuration by intervening clouds in the mixed case
\begin{equation}
\frac{I_0}{I_{\rm ext}}=\frac{\Lambda}{1-{\rm e}^{-\Lambda}}\ .
\end{equation}
Complete absorption $I_0/I_{\rm ext} \gg 1$ is only reached in the innermost
part of the AGN torus ($R \ll 1$~pc).

The screen and the mixed models can be regarded as two extreme cases and the reality maybe
somewhere in between these two models. Most probably the starburst will be 
significantly obscured at galactocentric radii smaller than $1$~pc. The extinction of star formation 
at the 10~pc scale by torus clouds is so low that additional extinction by extended dust lanes at 
larger radii ($>100$~pc) in the galaxies is possible. 
We can therefore conclude that our torus model is consistent with the small
observed extinction of the central starburst.

\section{Fueling the central engine \label{sec:fuel}}

The inner edge of the torus is thought to be set by the dust sublimation radius which is located at
about $0.1$-$0.5$~pc from the central black hole. In Sect.~\ref{sec:comparison}
we have derived the mass accretion rate of the tori, i.e. the mass transport rate arriving at these inner
edges. It is not clear what happens between the inner edge of the torus and
the thin accretion disk around the black hole and what the relation is to the maser emission regions.
These maser disks have sizes of about $0.1$~pc. Between $0.1$~pc and $0.5$~pc from the
central black hole an X-ray heated wind is most likely formed (Krolik \& Kriss 1995, 2001).
Less dense and sheared clouds lose their dust by evaporation at the sublimation temperature are consequently
ionized by the AGN X-ray emission, heated, and blown away in a line-driven wind. 
What remains of the dust or is newly formed in the wind to obscure the AGN at even higher inclination 
angles is undetermined in this scenario. The mass accretion rate onto the central black hole $\dot{M}_{\rm BH}$
is thus the difference between the mass accretion within the torus $\dot{M}$ and the 
mass loss due to the AGN wind $\dot{M}_{\rm wind}$:
\begin{equation}
\dot{M}_{\rm BH}=\dot{M}-\dot{M}_{\rm wind}\ .
\end{equation}
Typical wind mass loss rates are $\dot{M}_{\rm wind}=0.03$-$0.3$~M$_{\odot}$yr$^{-1}$
(Blustin et al. 2005, 2007). 

\begin{figure}
        \resizebox{\hsize}{!}{\includegraphics{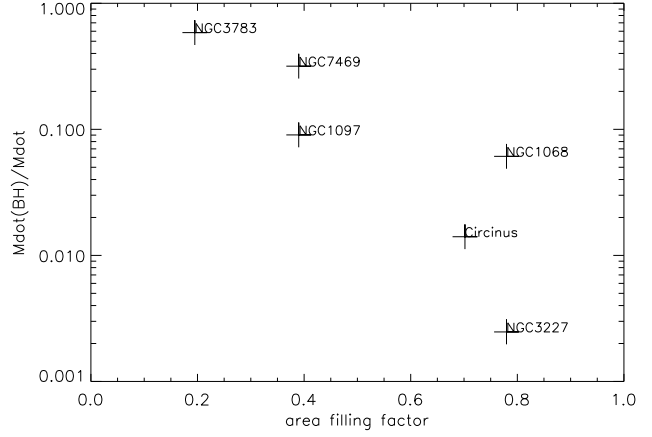}}
        \caption{Ratio between the mass accretion rate onto the central
	  black hole and the present torus mass accretion rate as a function
	  the area filling factor, i.e. $\Phi_{\rm A}=4/3/\vartheta$.
        } \label{fig:fuel}
\end{figure}

To investigate the relation between the mass accretion rate onto the central
black hole and the present torus mass accretion rate, we plot their ratio as a function
of the area filling factor $\Phi_{\rm A}=4/3\,\vartheta^{-1}$. 
The mass accretion rate onto the central black hole $\dot{M}_{\rm BH}$ is derived from the AGN luminosity
\begin{equation}
L=\eta \dot{M}_{\rm BH} c^2\ ,
\end{equation}
where the efficiency $\eta=0.1$ and $c$ is the speed of light.
The area filling factor is inversely proportional to the transparency of the torus (Fig.~\ref{fig:fuel}). 
Small area filling factors correspond to transparent tori.

We observe a tentative trend in the sense that in opaque tori only $0.1$-$10\,\%$ of the
torus mass accretion rate feeds the central black hole. On the other hand,
if the torus is transparent, $10$-$100\,\%$ of the torus mass accretion rate
is used to feed the central engine.

The main differences between opaque ($\vartheta \sim 1$) and transparent tori ($\vartheta > 3$)
are the cloud collision rate and the cloud mean free path. Since there is no
correlation with the collision rate $t_{\rm coll}=\vartheta/\Omega$, we suspect
the mean free path to be responsible for the accretion efficiency of the tori.
The different mean free paths between the clouds means that in transparent tori
the AGN emission can reach all clouds of the torus, whereas in the opaque case
it only reaches clouds at small galactocentric radii, i.e. close to the central
source. Due to this radiation only the densest clouds can survive, the less denser
clouds being evaporated by a photodissiciation or X-ray dissociation region (PDR/XDR).
Vollmer \& Duschl (2001) calculated the location of the ionization fronts in the
cloud of the circumnuclear disk in the Galactic Center. These clouds have about the
same mass as the AGN torus clouds ($M_{\rm cl} \sim 10$~M$_{\odot}$) and sizes of 
$\sim 0.1$~pc. Whereas these clouds are close to the shear limit, the AGN torus 
clouds are $\sim 100$ times denser.
Note however, that in the
Galactic Center the source of ionization is a central cluster of about 40 O stars
which is much weaker than an AGN.

Vollmer \& Duschl (2001) showed that the cloud radius due to the ionization front
is given by
\begin{equation}
r_{\rm cl}=3.645\,10^{15} J_{\rm UV}^{-\frac{1}{3}} c_{\rm i}^{-\frac{4}{3}} c_{\rm s}^{\frac{8}{3}}\ ,
\end{equation}
where $J_{\rm UV}$ is the number of UV photons per cm$^{2}$ and sec, $c_{\rm i}$ is the
sound speed in the ionized gas, and $c_{\rm s}$ is the sound speed in the cloud.
If one assumes that the cloud temperature is determined by the external radiation
field $c_{\rm s} \propto T^{\frac{1}{2}} \propto J_{\rm UV}^{\frac{1}{8}}$, the cloud radius
does not change with the cloud's distance from the galactic center.

We scale up the Galactic Center by a factor of $10^4$ and obtain
$c_{\rm i}=7.6$~km\,s$^{-1}$, $c_{\rm s}=3$~km\,s$^{-1}$(R/(1~pc))$^{-\frac{1}{4}}$, and
$J_{\rm UV}=1.8\,10^{16}$~cm$^{-2}$s$^{-1}$(R/(1~pc))$^{-2}$ for a transparent torus.
This leads to a cloud radius of
$r_{\rm cl}=0.03$~pc, which is very close to the cloud radius obtained for
our sample of AGN tori (Table~\ref{tab:prestorus}). Thus, only dense, selfgravitating
clouds can survive in an illuminated environment, i.e. a transparent torus.
On the other hand, if the torus is opaque, less dense and larger clouds can survive.
Since we derive only the properties of the densest and most massive clouds in
our model which determine the physics of the outer torus, less dense and less massive clouds
are still consistent with the model as long as they do not dominate the disk mass. 
As a consequence, the transparency values derived for the individual AGNs 
and summarized in Table~\ref{tab:prestorus} are upper limits.
We therefore suggest that towards the inner edge of opaque tori clouds become
less dense with densities close to the shear limit.
These clouds, however, once they arrive at the inner edge of the torus, are (i) destroyed
easily by the influence of shear and possible winds and are (ii) much easier ionized and
evaporated by the AGN emission.
In addition, for AGN close to the Eddington limit H\"onig \& Beckert (2007) showed 
that dusty clouds experience a strong radiation pressure. For transparent tori only dense 
and compact clouds smaller than the shear limit will be found, while for opaque tori shadowing 
by clouds at the inner edge allows larger clouds to survive at intermediate distances.

This tentative picture leads to two predictions for AGN tori that might be verified in the future:
\begin{itemize}
\item
the inner parts of opaque tori should harbor larger clouds and should therefore have larger volume 
filling factors,
\item
the ratio of  the torus mass accretion rate to the mass removed in winds generated at the inner boundary 
of tori should be lower for opaque tori.
\end{itemize}

\section{Uncertainties}

Up to this point we ignored any possible observational errors.
In this section  the influence of the quantities derived
from observations on the correlations shown in Fig.~\ref{fig:tmassacc} and
Fig.~\ref{fig:fuel} is investigated. We assume the following uncertainties:
\begin{itemize}
\item
For the gas mass ($M_{\rm gas}$) - about a factor 2, because of the uncertainty in mass distribution and 
hence in interpreting the kinematics;
\item
peak star formation rate ($\dot{M}_*^{\rm peak}$) - at least a factor 2-3 since the extinction is not
well constrained;
\item
starburst age ($t_{\rm SB}$) - up to a factor 2, assuming that all diagnostics are 
explained by a single burst, which we reject for NGC3783;
\item
rotation velocity ($v_{\rm rot}$) - $\sim 40\,\%$, since it is the
equivalent Keplerian rotation velocity, and depends on both the
measured dispersion and inclination-corrected velocity.
\item
turbulent velocity dispersion ($v_{\rm turb}$) - $\sim 20\,\%$ assuming that the velocity 
dispersion is not strongly affected by a wind.
\end{itemize}
All uncertainties of our derived quantities are thus dominated by the observational
uncertainties on $M_{\rm gas}$ and $\dot{M}_*$.
For the torus mass accretion rate we have
\begin{equation}
\dot{M}=\frac{128}{\pi^4} f v_{\rm turb}^2 v_{\rm rot}^{-2} M_{\rm gas}^2 M_{\rm dyn}^{-2} c_{\rm s}^{-4} \Omega M_{\rm cl}\ .
\end{equation}
During the torus evolution the mass accretion rate is proportional to the square of
the gas mass $\dot{M} \propto M_{\rm gas}^2$.
The gas mass of the massive ($Q \sim 1$) turbulent disk only depends on the
peak star formation rate
\begin{equation}
M_{\rm gas} \propto (\dot{M}_*^{\rm peak})^{\frac{5}{11}}\ . 
\end{equation}
We took this gas mass as the initial gas mass of the collisional torus at the
beginning of Phase II.
Thus the ratio between the mass accretion rate of the present torus $\dot{M}$
and that of the initial collisional torus $\dot{M}_{\rm old}$ (Fig.~\ref{fig:tmassacc})
depends on the square of the present gas mass $M_{\rm gas}$
and on the inverse of the peak star formation rate $\dot{M}_*^{\rm peak}$:
\begin{equation}
\frac{\dot{M}}{\dot{M}_{\rm old}} \propto \frac{M_{\rm gas}^2}{\dot{M}_*^{\rm peak}}\ .
\end{equation}
On the other hand, the ratio between the mass accretion rate onto the black hole
$\dot{M}_{\rm BH}$ and the torus mass accretion rate $\dot{M}$ depend on the
AGN luminosity $L$ and on the inverse of the square of the gas mass $M_{\rm gas}$:
\begin{equation}
\frac{\dot{M}_{\rm BH}}{\dot{M}} \propto \frac{L}{M_{\rm gas}^2}\ .
\end{equation}
An underestimation of the gas mass, which we assumed to be $10\,\%$ of the dynamical mass,
thus leads to a strong increase of the two ratios, while
an underestimation of the peak star formation rate and the AGN luminosity leads to a decrease of 
the two ratios. In extreme cases both ratios can have errors up to a factor of $10$.

The starburst ages have uncertainties of a factor of 2.
The area filling factor is proportional to the inverse of the gas mass
$\Phi_{\rm V} \propto M_{\rm gas}^{-1}$ (Eq.~\ref{eq:taupres}).
Thus, the associated uncertainty is also a factor of 2.

Despite these relatively large uncertainties we believe that the correlations shown in
Fig.~\ref{fig:tmassacc} and \ref{fig:fuel} are real. 
All systematic errors, like an overestimate of the gas mass, would alter but not destroy 
the correlations, as long as the errors for different targets are not random. However, to make
our findings more robust more spectroscopic VLT SINFONI observations with high spatial resolution of
AGNs are needed. Most importantly, future ALMA high resolution CO line observations
are necessary to determine the total gas mass of the tori with an
uncertainty of $\sim 10-30$\,\%.

\section{A holistic view of torus evolution in AGN \label{sec:holistic}}

VLT SINFONI observations of the close environments ($\sim 30$~pc) of a sample of nearby AGNs
by Davies et al. (2007) showed that thick gas tori and recent central starbursts 
with ages smaller than $100$~Myr are ubiquitous.
We compare different clumpy accretion disk models to these observations:
\begin{itemize}
\item
fully gravitational turbulent disks where the turbulence is maintained
by the energy input from the gravitational potential via mass accretion,
\item
supernova and stellar wind driven turbulent disks (SN model) where the turbulence is maintained by
stellar winds and supernova explosions,
\item
collisional disks where the orbital motion is randomized by partially elastic collisions 
which also allow mass transport to the center and angular momentum redistribution.
\end{itemize}
Whereas the measured rotation velocity, turbulent velocity dispersion, and gas mass
tell us about the current state of the gas torus, the peak star formation rate
and the age of the starburst provide information on the past appearance of the torus. 
We assume that the physical properties of the torus are mainly determined
by external mass accretion from scales of $\sim 100$~pc.

The result of this work is a time sequence for the torus evolution.
Present tori appear to be collisional and geometrically thick whereas the tori giving rise to the starburst
in the past are of turbulent nature and relatively thin.
The torus evolution can be divided into 3 phases depending on the external mass accretion rate:
\begin{itemize}
\item
Phase I: initial massive infall: formation of a massive turbulent stellar 
wind-driven gas disk with $Q \sim 1$,
\item
Phase II: decreasing, but still high mass accretion rate: collisional thick torus,
\item
Phase III: decreasing, now low mass accretion rate: collisional thin torus.
\end{itemize}
Phase I is short ($\sim 10$~Myr). Once the SN explode, they remove the intercloud medium and clear the disk
leaving behind only the densest clouds.
The result is a collisional torus.
All tori discussed in this paper 
are interpreted to be in phase II. Therefore this
phase can last for more than $100$~Myr. 
The transparency of these tori depend on their Toomre parameter, the turbulent
velocity, and the internal pressure of the clouds (temperature and magnetic fields). 
If there is a second short massive infall, the torus can again switch into phase I.
We suggest that this has happened in  
the case of NGC~3783. On the other hand, an initial
massive gas infall is not mandatory as  the case of Circinus illustrates.
Once the external mass accretion rate has significantly decreased, the torus
becomes thin and transparent. The Circumnuclear Disk in the Galactic Center
represents this phase.

In addition we conclude that 
\begin{enumerate}
\item
The massive turbulent stellar wind-driven gas disk (phase I) gives rise to a starburst
with a star formation rate that equals the disk mass accretion rate.
\item
The collisional torus also forms stars, but with an efficiency
which is about $10$\,\% that of the turbulent disk.
\item
During the evolution of the thick collisional torus (phase II) the decrease of the
mass accretion rate is slow (a factor of 4 within $150$~Myr). 
\item
The present collisional tori (phase II) do not significantly obscure the
stellar populations born during the recent starburst. This is the case even for
Sy2 nuclei.
\item
We find a tentative correlation between the area filling factor of the clouds
and the ratio between the mass accretion rate onto the central black hole
and the torus mass accretion rate. AGNs with a high area filling factor
lose more than $90$\,\% of the mass transported through the torus in a thermal torus wind
at the inner edge. We suggest that this is due to shadowing of the
gas clouds from the central AGN throughout most the extended torus which allows the existence of low
density clouds whose densities are close to the shear limit.
\end{enumerate}

Future VLT SINFONI and ALMA observations will be necessary to confirm
our proposed torus evolution sequence.



\begin{thebibliography}{}

\bibitem{a1} Antonucci R., 1993, ARA\&A, 31, 473

\bibitem{a2} Asari, N., Vega, L., Garcia-Rissmann, A., Gonz�lez Delgado, R., Storchi-Bergmann, T., \& Cid Fernandes, R. 2007, in IAU Symp. 235, Galaxy Evolution across the Hubble Time, ed. F. Combes \& J. Palous (Cambridge: Cambridge Univ. Press), 71

\bibitem{a3} Barvainis R., 1987, ApJ, 320, 537

\bibitem{t1} Beckert, T., \& Duschl, W.~J., 2004, A\&A, 426, 445

\bibitem{a4} Blustin, A.J., Page, M.J., F\"{u}rst, S.V. et al. 2005, A\&A, 431, 111

\bibitem{a5} Blustin, A.J., Kriss, G.A., Holczer, T. et al. 2007, A\&A, 466, 107

\bibitem{a6} Cid Fernandes, R., Gu, Q., Melnick, J., et al. 2004, MNRAS, 355, 273

\bibitem{a7} Davies, R.I., Thomas, J., Genzel, R. et al. 2006, ApJ, 646, 754

\bibitem{a8} Davies, R.I., Mueller-Sanchez, F., Genzel, R., et al. 2007, ApJ, 671, 1388

\bibitem{a9} Elvis M., Wilkes B., McDowell J. et al., 1994, ApJS, 95, 1

\bibitem{a9a} Erickson E.F., Colgan S.W.J., Simpson J.P., Rubin R.H., Hass M.R., 194, in: Genzel and Harris (1994), 249

\bibitem{a10} Gonz\'alez Delgado, R., \& Cid Fernandes, R. 2005, in Starbursts: From 30 Doradus to Lyman Break Galaxies, ed. R. de Grijs \& R. Gonz�lez Delgado (Dordrecht: Springer), 263

\bibitem{a11} G\"{u}sten R., Genzel R., Wright M.C.H. et al., 1987, ApJ 318, 124

\bibitem{a11a} Hicks E., Davies R., Malkan M., Genzel R., Tacconi L., Mueller S\'anchez F., Friedrich S., Sternberg A., 2008, ApJ, submitted

\bibitem{a12} H\"{o}nig, S.F., Beckert, T., Ohnaka, K., \& Weigelt, G. 2006, A\&A, 452, 459

\bibitem{t2} H\"{o}nig, S.F. \& Beckert, T., 2007, MNRAS, 380, 1172


\bibitem{a14} Krolik J.H. \& Begelman M.C. 1988, ApJ, 329, 702

\bibitem{a15} Krolik, J.H. \& Kriss, G.A. 1995, ApJ, 447, 512

\bibitem{a16} Krolik, J.H. \& Kriss, G.A. 2001, ApJ, 561, 684

\bibitem{a17} Mac Low, M.-M. \&  Klessen, R.S. 2004, RvMP, 76, 125


\bibitem{a18a} Plante, R.L., Lo, K.Y., \& Crutcher, R.M. 1995, ApJ, 445, L113

\bibitem{a18aa} Pringle, J.E. 1981, ARA\&A, 19, 137 

\bibitem{a18b} R\"{o}diger E. \& Hensler G. 2008, A\&A, 483, 121

\bibitem{a19} Sanders D. B., Phinney E. S., Neugebauer G., Soifer B. T., Matthews K., 1989, ApJ, 347, 29

\bibitem{a20} Sarzi, M., Shields, J., Pogge, R., \& Martini, P. 2007, in IAU Symp. 241, Stellar Populations as Building Blocks of Galaxies, ed. R. Peletier \& A. Vazdekis (Cambridge: Cambridge Univ. Press), 489

\bibitem{a21} Veilleux, S., Cecil, G., \& Bland-Howthorn, J. 2005, ARA\&A, 43, 769


\bibitem{a22} Vollmer B. \& Duschl W.J. 2001, A\&A, 377, 1016


\bibitem{a24} Vollmer B. \& Beckert T. 2002, A\&A, 382, 872, (Paper I)

\bibitem{a25} Vollmer B. \& Beckert T. 2003, A\&A, 404, 21, (Paper II)

\bibitem{a26} Vollmer B., Beckert, T., \& Duschl, W.J. 2004, A\&A, 413, 949


\end{thebibliography}
\end{document}